\documentclass[twocolumn,prb,aps,epsfig,superscriptaddress,showpacs]{revtex4-1}
\usepackage[latin9]{inputenc}
\usepackage{bm}
\usepackage{amsmath}
\usepackage{amssymb}
\usepackage{graphicx}
\usepackage{esint}
\usepackage[unicode=true,pdfusetitle,
 bookmarks=true,bookmarksnumbered=false,bookmarksopen=false,
 breaklinks=false,pdfborder={0 0 1},colorlinks=false]{hyperref}
\usepackage{bm}
\usepackage{amsfonts}
\usepackage{mathrsfs}
\usepackage{amsbsy}
\usepackage{epstopdf}

\hypersetup{
 colorlinks,citecolor=red}
\makeatletter
\@ifundefined{textcolor}{}
{
 \definecolor{BLACK}{gray}{0}
 \definecolor{WHITE}{gray}{1}
 \definecolor{RED}{rgb}{1,0,0}
 \definecolor{GREEN}{rgb}{0,1,0}
 \definecolor{BLUE}{rgb}{0,0,1}
 \definecolor{CYAN}{cmyk}{1,0,0,0}
 \definecolor{MAGENTA}{cmyk}{0,1,0,0}
 \definecolor{YELLOW}{cmyk}{0,0,1,0}
}
\makeatother

\begin{document}

\title{Hopf insulators and their topologically protected surface states}
\date{\today}
\author{D.-L. Deng$^{1,2}$, S.-T. Wang$^{1,2}$, C. Shen$^{1,2}$, and L.-M.
Duan}
\affiliation{Department of Physics, University of Michigan, Ann Arbor, Michigan 48109, USA}
\affiliation{Center for Quantum Information, IIIS, Tsinghua University, Beijing 100084,
PR China}

\begin{abstract}
Three-dimensional (3D) topological insulators in general need to be
protected by certain kinds of symmetries other than the presumed $U(1)$
charge conservation. A peculiar exception is the Hopf insulators which are
3D topological insulators characterized by an integer Hopf index. To
demonstrate the existence and physical relevance of the Hopf insulators, we
construct a class of tight-binding model Hamiltonians which realize all
kinds of Hopf insulators with arbitrary integer Hopf index. These Hopf
insulator phases have topologically protected surface states and we
numerically demonstrate the robustness of these topologically protected
states under general random perturbations without any symmetry other than
the $U(1)$ charge conservation that is implicit in all kinds of topological
insulators.
\end{abstract}

\maketitle

Topological phases of matter may be divided into two classes: the intrinsic
ones and the symmetry protected ones \cite{2013XieChen}. Symmetry protected
topological (SPT) phases are gapped quantum phases that are protected by
symmetries of the Hamiltonian and cannot be smoothly connected to the
trivial phases under perturbations that respect the same kind of symmetries.
Intrinsic topological (IT) phases, on the other hand, do not require
symmetry protection and are topologically stable under arbitrary
perturbations. Unlike SPT phases, IT phases may have exotic excitations
bearing fractional or even non-Abelian statistics in the bulk \cite%
{2008Nayak}. Fractional~\cite{FQHE} quantum Hall states and spin liquids~%
\cite{SpinLiquid} belong to these IT phases. Remarkable examples of the SPT
phases include the well known $2$D and $3$D topological insulators and
superconductors protected by time reversal symmetry \cite%
{2007Fu,2010Hassan,2011Qi}, and the Haldane phase of the spin-$1$ chain
protected by the $SO(3)$ spin rotational symmetry \cite{HaldanePhase}. For
interacting bosonic systems with on-site symmetry $G$, distinct SPT phases
can be systematically classified by group cohomology of $G$ \cite%
{2013XieChen}, while for free fermions, the SPT phases can be systematically
described by K-theory or homotopy group theory \cite{2003Nakahara}, which
leads to the well known periodic table for topological insulators and
superconductors \cite{2009Kitaev,2008Schnyder}.

Most 3D topological insulators have to be protected by some other symmetries
\cite{2008Schnyder,2009Kitaev}, such as time reversal, particle hole or
chrial symmetry, and the $U(1)$ charge conservation symmetry~\cite%
{2013Budich}. A peculiar exception occurs when the Hamiltonian has just two
effective bands. In this case, interesting topological phases, the so-called
Hopf insulators \cite{2008Moore}, may exist. These Hopf insulator phases
have no symmetry other than the prerequisite $U(1)$ charge conservation. To
elucidate why this happens, let us consider a generic band Hamiltonian in $3$%
D with $m$ filled bands and $n$ empty bands. Without symmetry constraint,
the space of such Hamiltonians is topologically equivalent to the
Grassmannian manifold $\mathbb{G}_{m,m+n}$ and can be classified by the
homotopy group of this Grassmannian \cite{2008Schnyder}. Since the homotopy
group $\pi _{3}(\mathbb{G}_{m,m+n})=\{0\}$ for all $(m,n)\neq (1,1)$, there
exists no nontrivial topological phase in general. However, when $m=n=1$, $%
\mathbb{G}_{1,2}$ is topologically equivalent to $\mathbb{S}^{2}$ and the
well-known Hopf map in mathematics shows that $\pi _{3}(\mathbb{G}%
_{1,2})=\pi _{3}(\mathbb{S}^{2})=\mathbb{Z}$ \cite{2003Nakahara}. This
explains why the Hopf insulators may exist only for Hamiltonians with two
effective bands. The classification theory shows that the peculiar Hopf
insulators may exist in $3$D, but it does not tell us which Hamiltonian can
realize such phases. It is even a valid question whether these phases can
appear at all in physically relevant Hamiltonians. Moore, Ran, and Wen made
a significant advance in this direction by constructing a Hamiltonian that
realizes a special Hopf insulator with the Hopf index $\chi =1$ \cite%
{2008Moore}.

In this Rapid Communication, we construct a class of tight-binding
Hamiltonians that realize arbitrary Hopf insulator phases with any integer
Hopf index $\chi $. The Hamiltonians depend on two parameters and contain
spin-dependent and spin-flip hopping terms. We map out the complete phase
diagram and show that all the Hopf insulators can be realized with this type
of Hamiltonian. We numerically calculate the surface states for these
Hamiltonians and show that they have zero energy modes that are
topologically protected and robust to arbitrary random perturbations with no
other than the $U(1)$ symmetry constraint.

To begin with, let us notice that any two-band Hamiltonian in $3$D with one
filled band can be expanded in the momentum space with three Pauli matrices $%
\bm{\sigma}=(\sigma ^{x},\sigma ^{y},\sigma ^{z})$ as
\begin{equation}
\mathcal{H}(\mathbf{k})=\mathbf{u}(\mathbf{k})\cdot \boldsymbol{\sigma },
\end{equation}%
where we have ignored the trivial energy-shifting term $u_0(\mathbf{k})%
\mathbf{I}_2$ with $\mathbf{I}_2$ being the $2\times 2$ identity matrix. By
diagonalizing $\mathcal{H}(\mathbf{k})$, we have the energy dispersion $E(%
\mathbf{k})=\pm |\mathbf{u}(\mathbf{k})|$, where $|\mathbf{u}(\mathbf{k})|=%
\sqrt{u_{x}^{2}(\mathbf{k})+u_{y}^{2}(\mathbf{k})+u_{z}^{2}(\mathbf{k})}$.
The Hamiltonian is gapped if $|\mathbf{u}(\mathbf{k})|>0$ for all $\mathbf{k}
$. For the convenience of discussion of topological properties, we denote $%
\mathbf{u(k)}=|\mathbf{u(k)}|(x\mathbf{(k)},y\mathbf{(k)},z\mathbf{(k)})$
with $x^{2}\mathbf{(k)}+y^{2}\mathbf{(k)}+z^{2}\mathbf{(k)}=1$.
Topologically, the Hamiltonian (1) can be considered as a map from the
momentum space $\mathbf{k=}\left( k_{x},k_{y},k_{z}\right) $ characterized
by the Brillouin zone $\mathbb{T}^{3}$ ($\mathbb{T}$ denotes a circle and $%
\mathbb{T}^{3}$ is the 3D\ torus) to the parameter space $\mathbf{u(k)}%
\propto (x\mathbf{(k)},y\mathbf{(k)},z\mathbf{(k)})$ characterized by the
Grassmannian $\mathbb{G}_{1,2}=\mathbb{S}^{2}$. Topologically distinct band
insulators correspond to different classes of maps from $\mathbb{T}%
^{3}\rightarrow \mathbb{S}^{2}$.

The classification of all the maps from $\mathbb{T}^{3}\rightarrow \mathbb{S}%
^{2}$ is related to the \textit{torus homotopy group} $\tau _{3}(\mathbb{S}%
^{2})$ \cite{1948Fox}. To construct non-trivial maps from $\mathbb{T}%
^{3}\rightarrow \mathbb{S}^{2}$, we take two steps, first from $\mathbb{S}%
^{3}\rightarrow \mathbb{S}^{2}$ and then from $\mathbb{T}^{3}\rightarrow
\mathbb{S}^{3}$. We make use of the following generalized Hopf map $f:%
\mathbb{S}^{3}\rightarrow \mathbb{S}^{2}$ known in the mathematical
literature \cite{1947Whitehead}
\begin{equation}
x+iy=2\lambda \eta _{\uparrow }^{p}\bar{\eta}_{\downarrow }^{q},\;z=\lambda
(|\eta _{\uparrow }|^{2p}-|\eta _{\downarrow }|^{2q}),  \label{eq:HopfMap}
\end{equation}%
where $p$, $q$ are integers prime to each other and $\eta _{\uparrow }$, $%
\eta _{\downarrow }$ are complex coordinates for $\mathbb{R}^{4}$ satisfying
$|\eta _{\uparrow }|^{2}+|\eta _{\downarrow }|^{2}=1$ with the normalization
$\lambda =1/(|\eta _{\uparrow }|^{2p}+|\eta _{\downarrow }|^{2q})$. Equation
(2) maps the coordinates $\left( \text{Re}[\eta _{\uparrow }],\text{Im}[\eta
_{\uparrow }],\text{Re}[\eta _{\downarrow }],\text{Im}[\eta _{\downarrow
}]\right) $ of $\mathbb{S}^{3}$ to the coordinates $(x,y,z)$ of $\mathbb{S}%
^{2}$ with $x^{2}+y^{2}+z^{2}=1$. The Hopf index for the map $f$ is known to
be $\pm pq$ with the sign determined by the orientation of $\mathbb{S}^{3}$
\cite{1947Whitehead}. We then construct another map $g:\mathbb{T}%
^{3}\rightarrow \mathbb{S}^{3}$ (up to a normalization), defined by the equation
\begin{eqnarray}
\eta _{\uparrow }\mathbf{(k)} &=&\sin k_{x}+it\sin k_{y},  \notag \\
\eta _{\downarrow }\mathbf{(k)} &=&\sin k_{z}+i(\cos k_{x}+\cos k_{y}+\cos
k_{z}+h),  \label{eq:TorusToS3}
\end{eqnarray}%
where $t$ and $h$ are constant parameters. The composite map $f\circ g$ from
$\mathbb{T}^{3}\rightarrow \mathbb{S}^{2}$ then defines the parameters $%
\mathbf{u(k)}\propto (x\mathbf{(k)},y\mathbf{(k)},z\mathbf{(k)})$ in the
Hamiltonian as a function of the momentum $\mathbf{k}$. From Eqs. (2) and
(3), we have $\mathbf{u(k)}=|\mathbf{u(k)}|(x\mathbf{(k)},y\mathbf{(k)},z%
\mathbf{(k)})=(\text{Re}[2\eta _{\uparrow }^{p}\bar{\eta}_{\downarrow }^{q}],%
\text{Im}[2\eta _{\uparrow }^{p}\bar{\eta}_{\downarrow }^{q}],[|\eta
_{\uparrow }|^{2p}-|\eta _{\downarrow }|^{2q}])$, with $|\mathbf{u(k)}|=%
\frac{1}{\lambda \mathbf{(k)}}$. The Hamiltonian $\mathcal{H}(\mathbf{k})=%
\mathbf{u}(\mathbf{k})\cdot \boldsymbol{\sigma }$ is $\left( p+q\right) $th
order polynomials of $\sin \left( \mathbf{k}\right) $ and $\cos \left(
\mathbf{k}\right) $, which corresponds to a tight-binding model when
expressed in the real space. The Hamiltonian contains spin-orbital coupling
with spin-dependent hopping terms. When we choose $p=q=1$ and $(t,h)=(1,-3/2)
$, the Hamiltonian (1) reduces to the special case studied in Ref.\ \cite%
{2008Moore}.

\begin{figure}[tbp]
\includegraphics[width=0.49\textwidth]{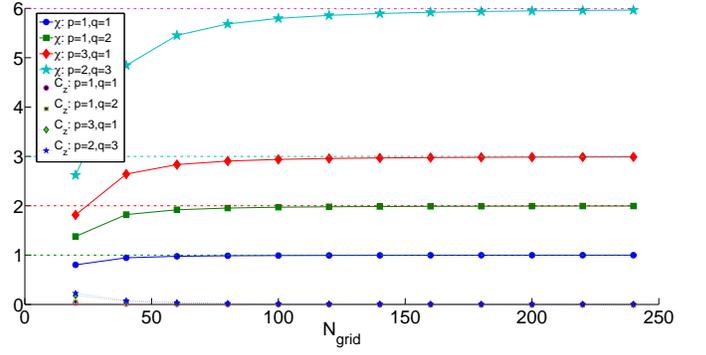}
\caption{(Color online) Plot of the Hopf index and the Chern number in
the $z$ direction for different $(p,q)$. The Hopf index and the Chern number
converge rapidly as the number of grids increases in discretization. The
parameters $t$ and $h$ are chosen as $(t,h)=(1,1.5).$ }
\label{fig:Numerical HopfI}
\end{figure}

\begin{figure}[tbp]
\includegraphics[width=0.47\textwidth]{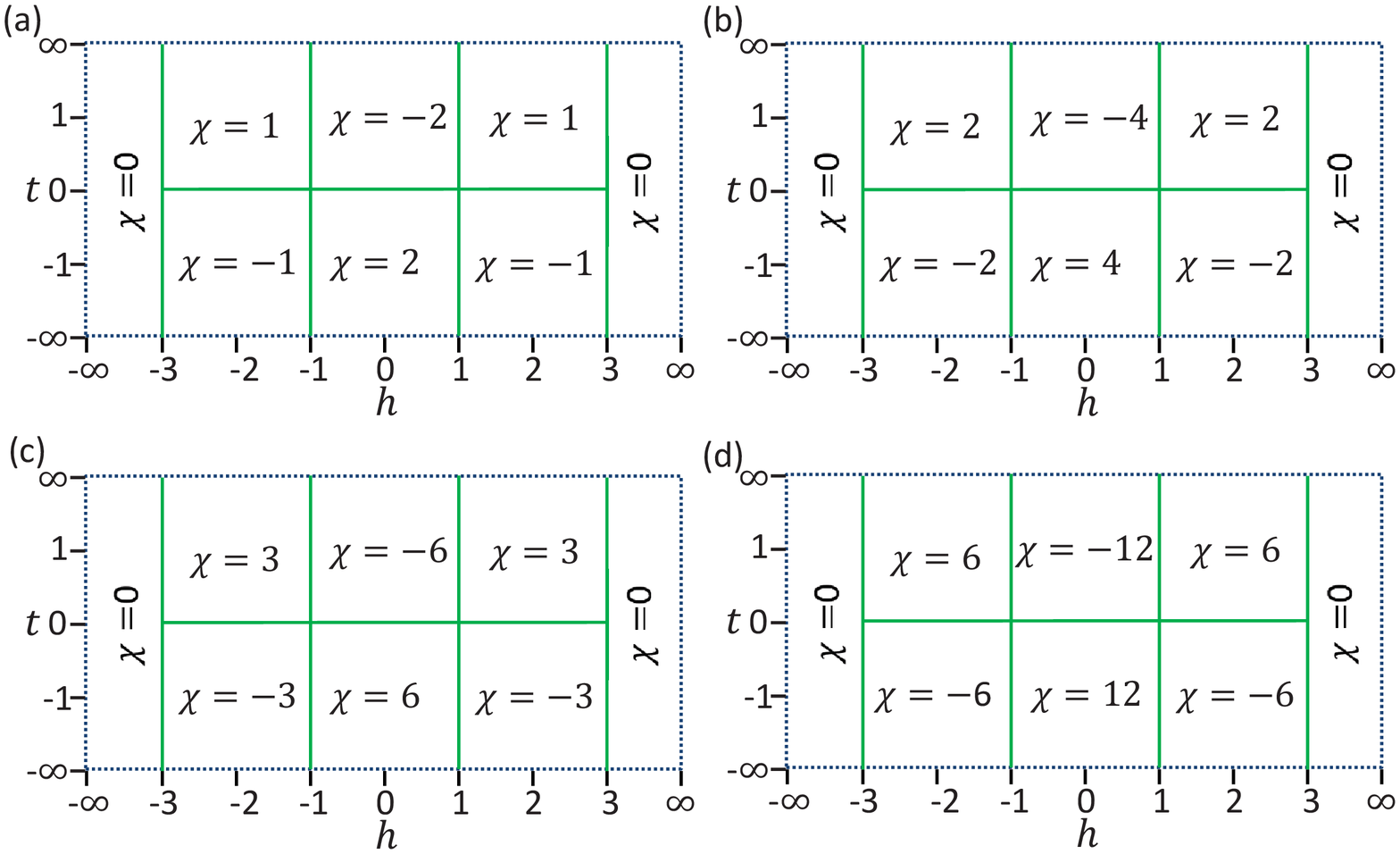}
\caption{(Color online) Phase diagrams of the Hamiltonian for different $%
(p,q)$. The values of $(p,q)$ in (a), (b), (c), and (d) are chosen to be $%
(1,1)$, $(1,2)$, $(3,1)$, and $(2,3)$, respectively.}
\label{fig:Phase Diagram}
\end{figure}

When the Hamiltonian is gapped with $|\mathbf{u}(\mathbf{k})|>0$, one can
define a direction on the unit sphere $\hat{\mathbf{u}}(\mathbf{k})=\left(
u_{x}(\mathbf{k}),u_{y}(\mathbf{k}),u_{z}(\mathbf{k})\right) /|\mathbf{u}(%
\mathbf{k})|=(x\mathbf{(k)},y\mathbf{(k)},z\mathbf{(k)})$. From $\hat{%
\mathbf{u}}(\mathbf{k})$, we define the Berry curvature\ $F_{\mu }=\frac{1}{%
8\pi }\epsilon _{\mu \nu \tau }\hat{\mathbf{u}}\cdot (\partial _{\nu }%
\mathbf{\hat{\mathbf{u}}\times \partial _{\tau }\hat{\mathbf{u}}})$, where $%
\epsilon _{\mu \nu \tau }$ is the Levi-Civita symbol and a summation over
the same indices is implied. A $3$D torus $\mathbb{T}^{3}$ has three
orthogonal cross sections perpendicular to the axis $x,y,z$, respectively.
For each cross section of space $\mathbb{T}^{2}$, one can introduce a Chern
number $C_{\mu }=\int_{-\pi }^{\pi }\int_{-\pi }^{\pi }dk_{\rho }dk_{\lambda
}F_{\mu }$, where $\mu =x,y,z$ and $\rho ,\lambda $ denote directions
orthogonal to $\mu $. To classify the maps from $\mathbb{T}^{3}\rightarrow
\mathbb{S}^{2}$ represented by $\hat{\mathbf{u}}(\mathbf{k})$, a topological
index, the so-called Hopf index, was introduced by Pontryagin \cite%
{1941Pontryagin}, who showed that the Hopf index takes values in the finite
group $\mathbb{Z}_{2\cdot \text{GCD}(C_{x},C_{y},C_{z})} $ when the Chern
numbers $C_{\mu }$ are nonzero \cite{1941Pontryagin}, where GCD denotes the
greatest common divisor. If the Chern numbers $C_{\mu }=0$ in all three
directions, the Hopf index takes all integer values $\mathbb{Z}$ and has a
simple integral expression \cite{1947Whitehead,1983Wilczek}
\begin{equation}
\chi \left( \hat{\mathbf{u}}\right) =-\int_{\text{BZ}}\mathbf{F\cdot \mathbf{%
A}}\;d\mathbf{k,}  \label{eq:Hopf-Index-1}
\end{equation}%
where $\mathbf{A}$ is the Berry connection (or called the gauge field) which
satisfies $\nabla \times \mathbf{A}=\mathbf{F}$. The Hopf index $\chi \left(
\hat{\mathbf{u}}\right) $ is gauge invariant although its expression depends
on $\mathbf{A}$. As we will analytically prove in the Appendix, the Chern
numbers $C_{\mu }=0$ for the map $\hat{\mathbf{u}}(\mathbf{k})$ defined
above in this paper in the gapped phase, so we can use the integral
expression of Eq. (4) to calculate the Hopf index $\chi \left( \hat{\mathbf{u%
}}\right) $. The index $\chi \left( \hat{\mathbf{u}}\right) $ can be
calculated numerically through discretization of the torus $\mathbb{T}^{3}$\
\cite{2008Moore}. Using this method, we have numerically computed the Hopf
index $\chi \left( \hat{\mathbf{u}}\right) $ for the Hamiltonian $\mathcal{H}%
(\mathbf{k})$ with various $p$ and $q$, and the results are shown in Fig.\ %
\ref{fig:Numerical HopfI}. As the grid number increases in discretization,
we see that the Chern numbers quickly drop to zero and the Hopf index
approaches the integer values $\pm pq$ or $\pm 2pq$ depending on the
parameters $t,h$. Based on the numerical results of $\chi \left( \hat{%
\mathbf{u}}\right) $, we construct the phase diagrams of the Hamiltonian (1)
for various $p$, $q$ in Fig.\ \ref{fig:Phase Diagram}. The phase boundaries
are determined from the gapless condition. The phase diagrams exhibit
regular patterns: they are mirror symmetric with respect to the axis $h=0$
and anti-symmetric with respect to the axis $t=0$. When $|h|>3$, we only
have a topologically trivial phase with $\chi \left( \hat{\mathbf{u}}\right)
=0$. From the result, we see that $\chi \left( \hat{\mathbf{u}}\right) $ has
an analytic expression with $\chi \left( \hat{\mathbf{u}}\right) =\pm pq$
when $1<|h|<3$ and $\chi \left( \hat{\mathbf{u}}\right) =\pm 2pq$ when $|h|<1
$.

To understand this result, we note that $\hat{\mathbf{u}}(\mathbf{k})$ is a
composition of two maps $\hat{\mathbf{u}}(\mathbf{k})=f\circ g(\mathbf{k})$.
The generalized Hopf maps $f$ from $\mathbb{S}^{3}\rightarrow \mathbb{S}^{2}$
has a known Hopf index $\pm pq$ \cite{1947Whitehead}. The maps $g$ from $%
\mathbb{T}^{3}\rightarrow \mathbb{S}^{3}$ can be classified by the torus
homotopy group $\tau _{3}(\mathbb{S}^{3})$ and a topological invariant has
been introduced to describe this classification \cite{2012Neupert}, which
has an integral expression
\begin{equation*}
\Gamma(g) =\frac{1}{12\pi ^{2}}\int_{\text{BZ}}d\mathbf{k}\epsilon _{\alpha
\beta \gamma \rho }\epsilon _{\mu \nu \tau }\frac{1}{|\bm{\eta}|^{4}}%
\bm{\eta}_{\alpha }\partial _{\mu }\bm{\eta}_{\beta }\partial _{\nu }%
\bm{\eta}_{\gamma }\partial _{\tau }\bm{\eta}_{\rho },\quad
\end{equation*}%
where $\bm{\eta}=(\text{Re}[\eta _{\uparrow }],\text{Im}[\eta _{\uparrow }],%
\text{Re}[\eta _{\downarrow }],\text{Im}[\eta _{\downarrow }])$. Direct
calculation of $\Gamma (g)$ leads to the following result:
\begin{equation*}
\Gamma (g)=%
\begin{cases}
0, & |h|>3 \\
1, & 1<|h|<3\;\text{and }t>0 \\
-2, & |h|<1\;\text{and }t>0.%
\end{cases}%
\end{equation*}%
Consequently, we have $\chi (\hat{\mathbf{u}})=\Gamma (g)\chi (f)=\pm
pq\Gamma (g)$, which is exactly the result shown in the phase diagrams in
Fig.\ \ref{fig:Phase Diagram}. A geometric interpretation is that $\Gamma
(g) $ counts how many times $\mathbb{T}^{3}$ wraps around $\mathbb{S}^{3}$
under the map $g$, and $\chi (f)$ describes how many times $\mathbb{S}^{3}$
wraps around $\mathbb{S}^{2}$ under the generalized Hopf map $f$. Their
composition gives the Hopf index $\chi (\hat{\mathbf{u}})$. A sign flip of $%
t $ changes the orientation of the sphere $\mathbb{S}^{3}$, which induces a
sign flip in $\chi (\hat{\mathbf{u}})$ and produces the anti-symmetric phase
diagram with respect to the axis $t=0$. As $(p,q)$ are arbitrary coprime
integers, $\chi (\hat{\mathbf{u}})$ apparently can take any integer value
depending on the values of $p,q$ and $t,h$. As a consequence, the
Hamiltonian $\mathcal{H}(\mathbf{k})$ constructed in this communication can realize
arbitrary Hopf insulator phases.

\begin{figure}[tbp]
\hspace{-8.5cm}{\scriptsize (a)} \newline
\hspace{-.4cm}\includegraphics[width=0.5\textwidth]{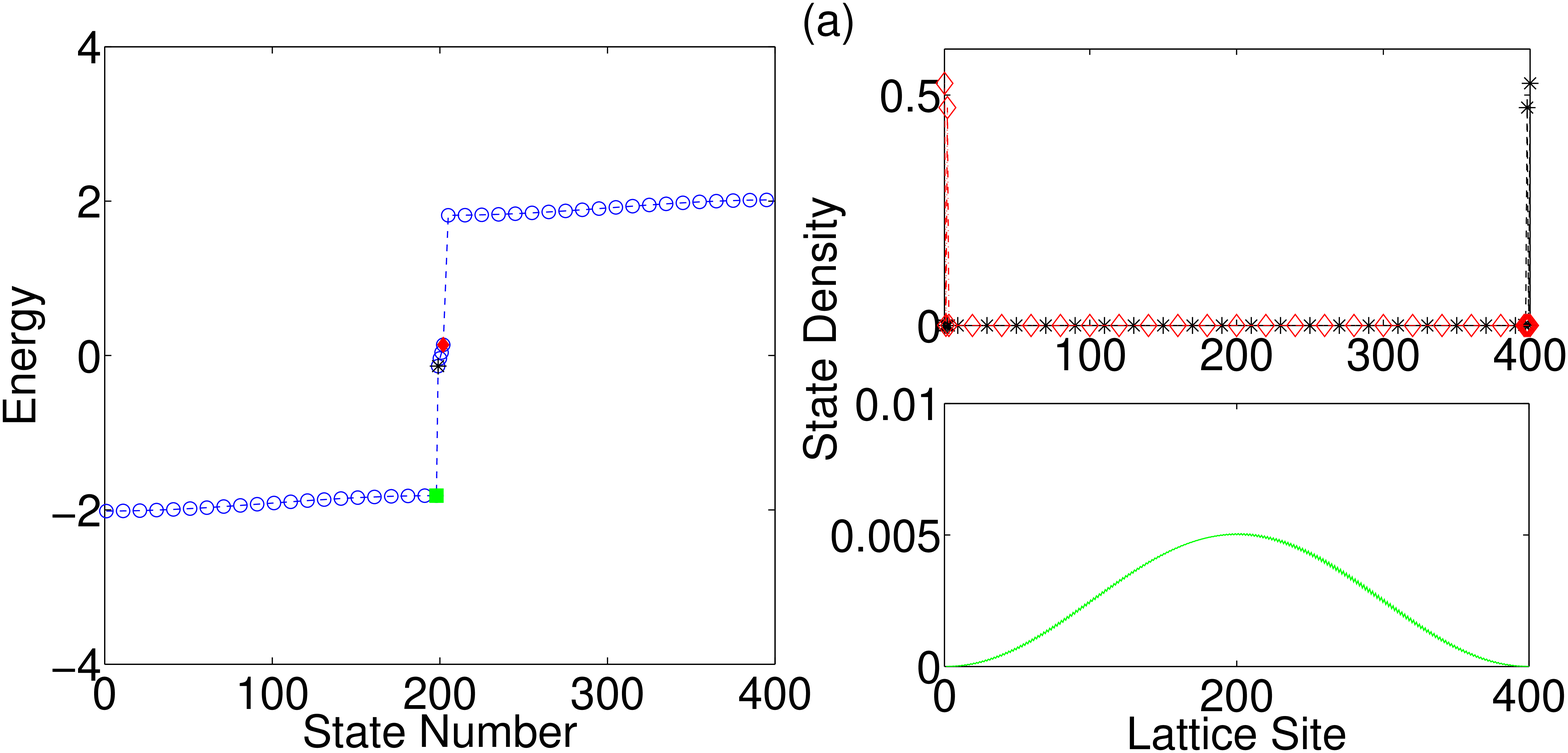}\newline
\vspace{-.85cm} \hspace{-8.5cm}{\scriptsize (b)} \newline
\hspace{-.4cm}\includegraphics[width=0.5\textwidth]{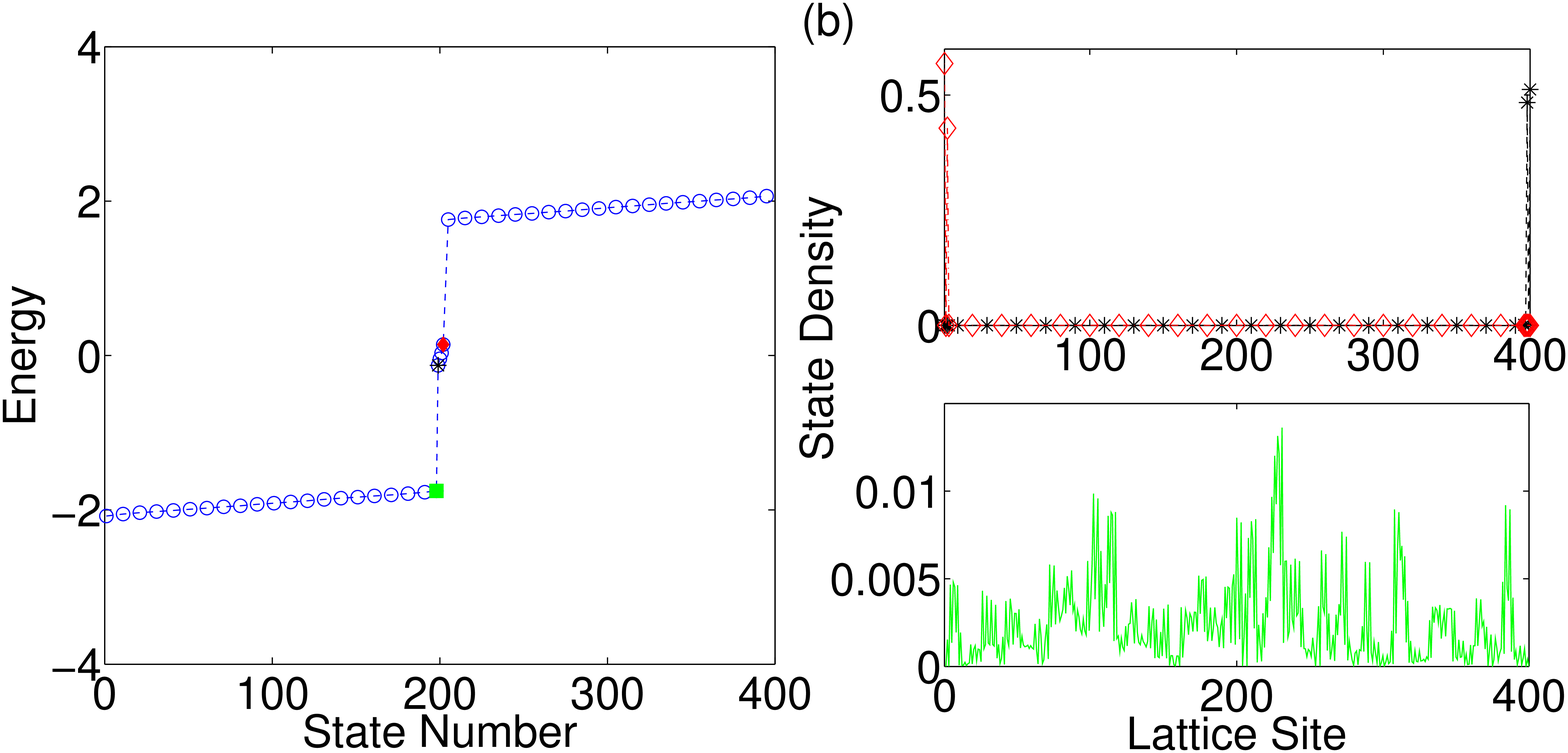}\newline
\vspace{-.85cm} \hspace{-8.5cm}{\scriptsize (c)} \newline
\hspace{-.4cm}\includegraphics[width=0.5\textwidth]{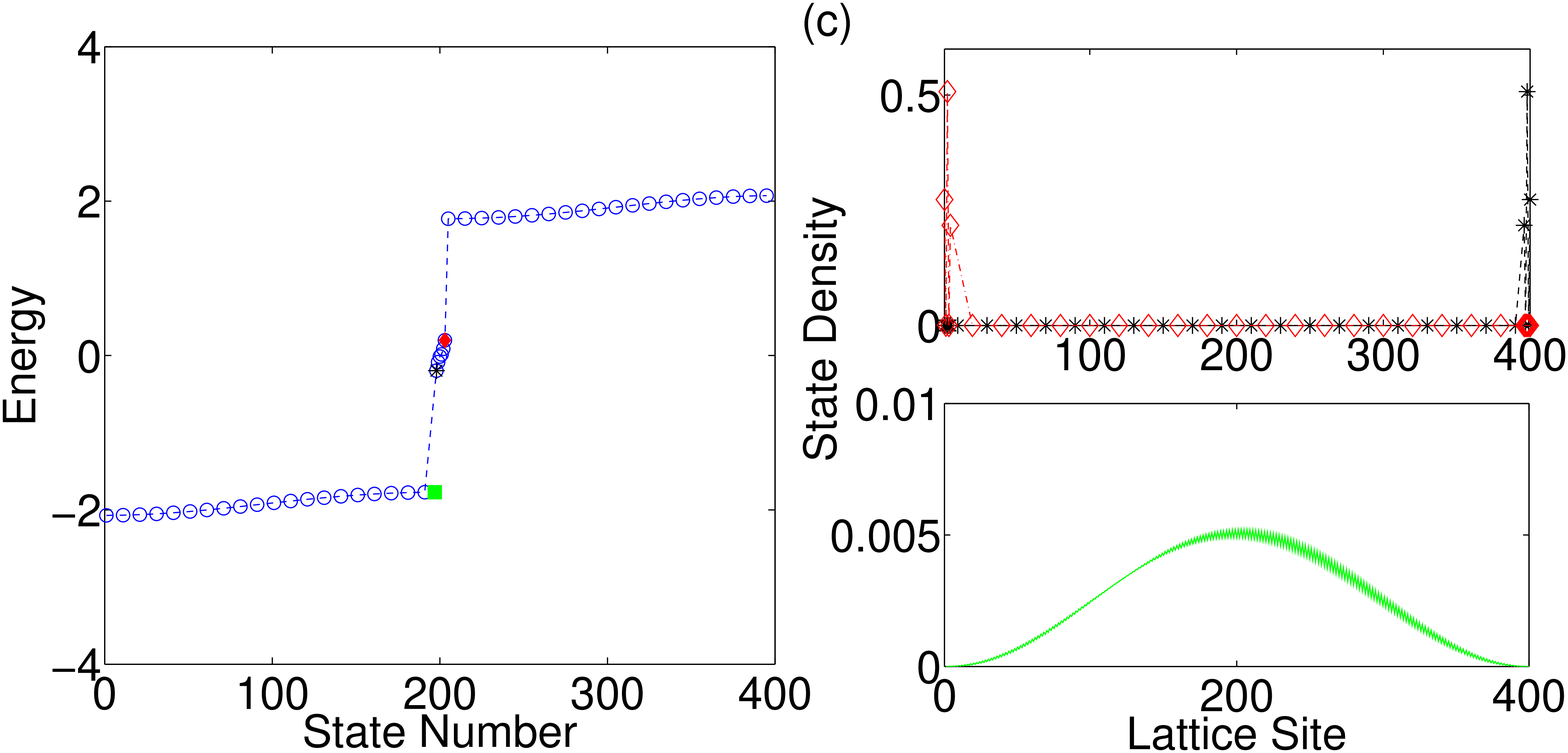}\newline
\vspace{-.85cm} \hspace{-8.5cm}{\scriptsize (d)} \newline
\hspace{-.4cm}\includegraphics[width=0.5\textwidth]{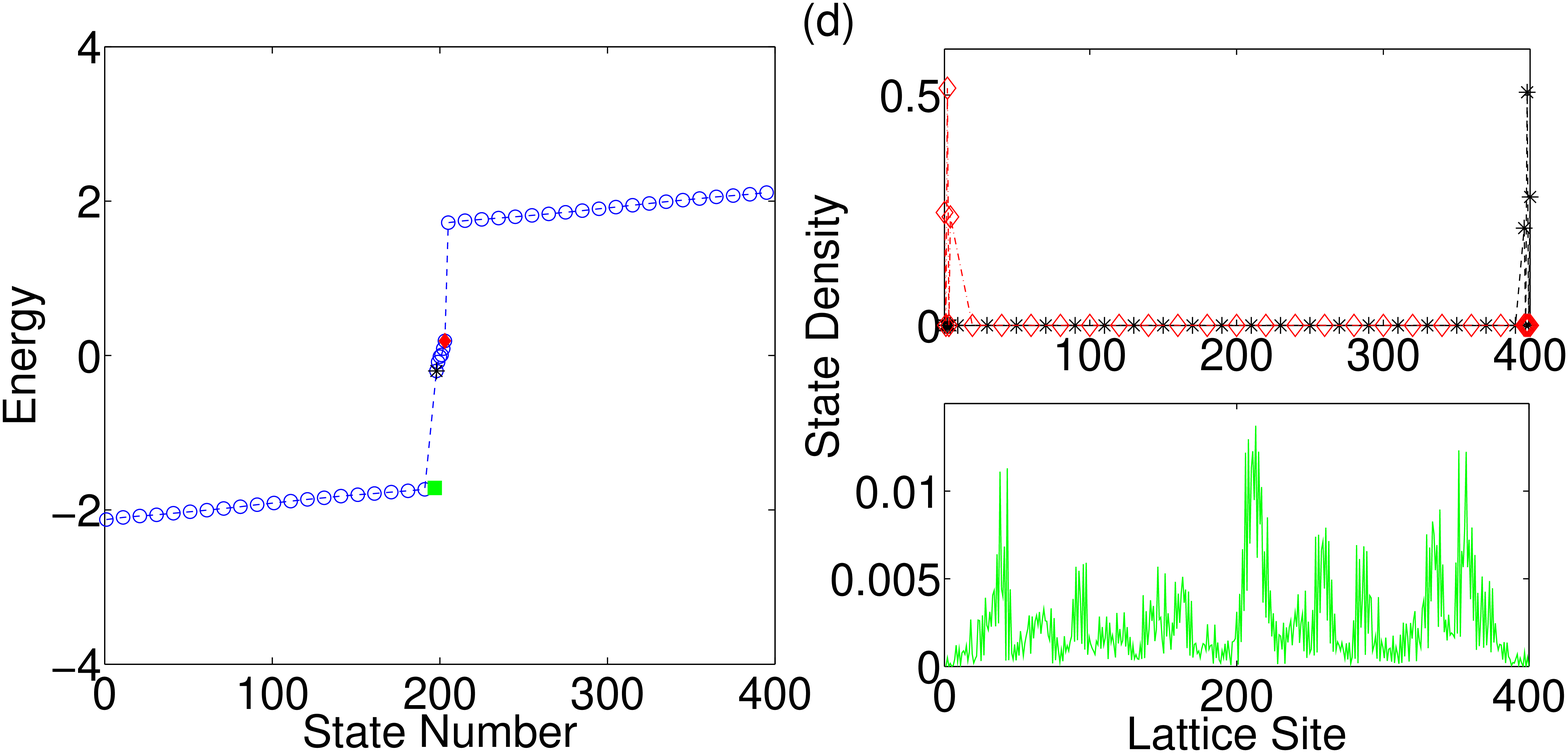} \vspace{%
-1cm}
\caption{(Color online) Surface states and zero-energy modes in the $(001)$
direction for a $200$-site-thick slab. The parameters $t$ and $h$ are chosen
as $(t,h)=(1,1.5)$ for all the figures. We have $(p,q)=(1,2)$ for (a,b) and $%
(p,q)=(1,3)$ for (c,d). In Fig. (b,d), we add random perturbations to the
Hamiltonian, but otherwise keep the same parameters as (a,c). The left
diagrams in (a,b,c,d) plot the energy spectrum of all $400$ states at a
fixed $(k_{x},k_{y})=(0.72,0.72)$ for easy visualization. The points inside
the gap represent the energies of the surface states. There are four (six)
surface states in (a,b) ((c,d)), respectively. The right diagrams in in
(a,b,c,d) show the wave functions of a surface state (upper one) and a bulk
state (lower one).}
\label{fig:SurfaceEdgeS}
\end{figure}

The nontrivial topological invariant guarantees existence of gapless surface
states at a smooth (i.e., adiabatic) boundary between a Hopf insulator and a
trivial insulator (or vacuum). Numerically, we find that gapless surface states are
still present even for sharp boundaries \cite{supplement1},
although we do not have an intuitive explanation why this is necessarily so as
the number of bands is not well-defined at a sharp boundary and the two-band condition
required for existence of the Hopf insulator could be violated at the surface.
Our results are summarized in Fig.\ \ref{fig:SurfaceEdgeS}.
From the figure, surface states and localized zero-energy modes are
prominent. These surface states are topologically protected and robust under
arbitrary random perturbations that only respect the prerequisite $U(1)$
symmetry. This can be clearly seen from Fig. \ref{fig:SurfaceEdgeS}: while
the wave functions of the bulk states change dramatically under random
perturbations, the wave functions of the surface states remain stable and
are always sharply peaked at the boundary. This verifies that the Hopf
insulators are indeed $3$D topological phases. Besides the results shown in
Fig. \ref{fig:SurfaceEdgeS}, we have calculated the surface states for a
number of different choices of parameters $(p,q)$ and $(t,h)$, and the
results consistently demonstrate that the surface states and zero energy
modes are always present and robust even to substantial perturbations unless
the bulk gap closes. Moreover, we roughly have more surfaces states when the
absolute value of the Hopf index becomes larger. However, this is not always
true. A direct correspondence between the Hopf index and the total winding
number of surface states may exist and deserves to be further investigated
\cite{2012Neupert}. It is also worthwhile to mention that these surface
states are extended/metallic in a clean crystal, as discussed in Ref. \cite%
{2008Moore}, but how disorder will affect these states is an important topic
that deserves further studies. The surface states might not be metallic with
disorder since there is no obvious way to protect these surface state from
localization without adding symmetries such as time-reversal.

An important and intriguing question is how to realize these Hopf insulators
in experiments. Laser assisted hopping of ultracold atoms in an optical
lattice offers a powerful tool to engineer various kinds of spin-dependent
tunneling terms \cite{AtomsInOL}, and thus provides a good candidate for
their realizations although the details still need to be worked out. Dipole
interaction between polar molecules in optical lattices also offers
possibilities to realize effective spin-dependent hopping \cite{2006Micheli}%
. As argued in Ref.\ \cite{2008Moore}, frustrated magnetic compounds such as
$\text{X}_{2}\text{Mo}_{2}\text{O}_{7}$ with $\text{X}$ being a rare earth
ion are other potential candidates. In addition, Hopf insulators may be
realized in 3D quantum walks\cite{2009Karski,2012Kitagawa}, where various
hopping terms are implemented by varying the walking distance and direction
in each spin-dependent translation and the robust surface states can be
observed with split-step schemes\cite{2012Kitagawa}.

In conclusion, we have introduced a class of tight-binding Hamiltonians that
realize arbitrary Hopf insulators. The topologically protected surface
states and zero-energy modes in these exotic phases are robust to random
perturbations that only respect the $U(1)$ charge conservation symmetry.
They are $3$D topological phases and sit outside of the periodic table \cite%
{2009Kitaev,2008Schnyder} for topological insulators and superconductors.

\textbf{Appendix. }Here, we prove that the Chern numbers $C_{\mu }=0$ in all
three directions for our Hamiltonian. Let us first consider $C_{x}$. To
prove $C_{x}=\int_{-\pi }^{\pi }\int_{-\pi }^{\pi
}dk_{y}dk_{z}F_{x}(k_{y},k_{z})=0$, it is sufficient to show $%
F_{x}(k_{y},k_{z})=-F_{x}(-k_{y},-k_{z})$, i.e., the function $F_{x}$ has an
odd parity under the exchange $(k_{y},k_{z})\rightarrow (-k_{y},-k_{z})$. We
denote the parity of a given function $\mathcal{F}(k_{y},k_{z})$ as $P[%
\mathcal{F}]=\{1,-1\}$ corresponding to $\left\{ \text{even, odd}\right\} $
parity. Our aim is to prove $P[F_{x}]=-1$. We let $g_{1}=\text{Re}(\eta
_{\uparrow }\mathbf{(k)})=\sin k_{x}$, $g_{2}=\text{Im}(\eta _{\uparrow }%
\mathbf{(k)})=t\sin k_{y}$, $g_{3}=\text{Re}(\eta _{\downarrow }\mathbf{(k)}%
)=\sin k_{z}$, and $g_{4}=\text{Im}(\eta _{\downarrow }\mathbf{(k)})=(\cos
k_{x}+\cos k_{y}+\cos k_{z}+h).$ Apparently, $P[g_{1}]=P[g_{4}]=1$ and $%
P[g_{2}]=P[g_{3}]=-1$ . We can normalize the $\mathbf{g}$-vector as $\hat{%
\mathbf{g}}=\mathbf{g}/|\mathbf{g}|=(g_{1},g_{2},g_{3},g_{4})/\sqrt{%
g_{1}^{2}+g_{2}^{2}+g_{3}^{2}+g_{4}^{2}}$. The components of $\hat{\mathbf{g}%
}$ have the same parity as the unnormalized ones. From the definition, we
have $\hat{u}_{x}=$Re$\left[ 2\hat{\lambda}(\hat{g}_{1}+i\hat{g}_{2})^{p}(%
\hat{g}_{3}-i\hat{g}_{4})^{q}\right] =2\hat{\lambda}$Re$\left[ \sum_{\alpha
=0}^{p}\sum_{\beta =0}^{q}C_{\alpha }^{p}C_{\beta }^{q}(-1)^{q-\beta
}i{}^{p+q-\alpha -\beta }\hat{g}_{1}^{\alpha }\hat{g}_{4}^{q-\beta }\hat{g}%
_{2}^{p-\alpha }\hat{g}_{3}^{\beta }\right] $, where $C_{\alpha }^{p}$ ($%
C_{\beta }^{q}$) denote the binormial coefficients and $\hat{\lambda}\equiv
1/(|(\hat{g}_{1}+i\hat{g}_{2})|^{2p}+|(\hat{g}_{3}+i\hat{g}_{4})|^{2q})$.
The exponent $p+q-\alpha -\beta $ of $i$ in $\hat{u}_{x}$ has to be even to
have a nonzero real part, so $P[\hat{u}_{x}]=P[\hat{g}_{2}^{p-\alpha }\hat{g}%
_{3}^{\beta }]=P[\hat{g}_{2}^{q-\beta }\hat{g}_{3}^{\beta }]=(-1)^{q}$.
Similarly, by using $\hat{u}_{y}=\text{Im}[2\hat{\lambda}(\hat{g}_{1}+i\hat{g%
}_{2})^{p}(\hat{g}_{3}-i\hat{g}_{4})^{q}]$, we find $P[\hat{u}_{y}]=-P[\hat{u%
}_{x}]$. Finally, from $\hat{u}_{z}=\hat{\lambda}(|(\hat{g}_{1}+i\hat{g}%
_{2})|^{2p}-|(\hat{g}_{3}+i\hat{g}_{4})|^{2q})$ we obtain $P[\hat{u}_{z}]=1$%
. As a consequence, $P[\hat{\mathbf{u}}\cdot (\partial _{\nu }\mathbf{\hat{%
\mathbf{u}}\times \partial _{\tau }\hat{\mathbf{u}}})]=-1$. Therefore, $%
P[F_{x}]=P[\hat{\mathbf{u}}\cdot (\partial _{k_{y}}\mathbf{\hat{\mathbf{u}}}%
\times \partial _{k_{z}}\hat{\mathbf{u}})]=-1.$This proves that $C_{x}=0$.
By the same parity arguments, we can show $C_{y}=C_{z}=0$.

\begin{acknowledgments}
We thank J. E. Moore, D. Thurston, K. Sun and X. Chen for helpful
discussions and J. Moore in particular for providing us his previous codes
for the calculation of the Hopf index. This work was supported by the NBR-
PC (973 Program) 2011CBA00300 (2011CBA00302), the DARPA\ OLE\ program, the
IARPA MUSIQC program, the ARO and the AFOSR MURI program.
\end{acknowledgments}

\onecolumngrid
\appendix
\clearpage
\section*{Supplemental Material: Hopf Insulators and Their Topologically
Protected Surface States}
\begin{quote}
In this supplemental material, we explain the details on how to obtain
the surface states and the zero energy modes.
\end{quote}

\begin{figure}[!t]
\includegraphics[width=0.95\textwidth]{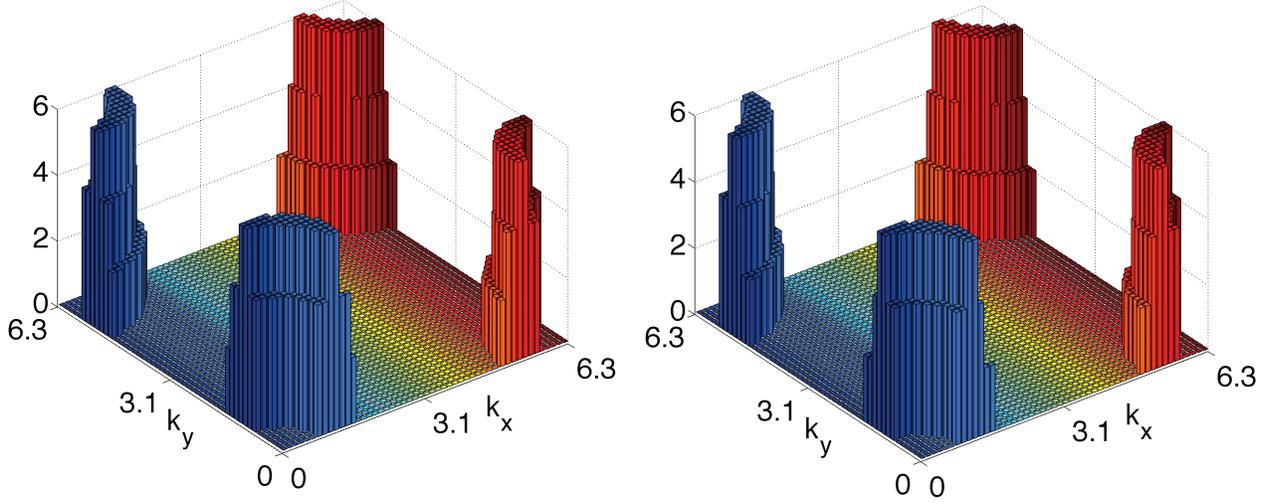}
\caption{(Color online) Number of surface states for each $k_{x}$ and $k_{y}$
in the (001) direction. Both diagrams show the case when $(p,q)=(1,3)$ and $(t,h)=(1,1.5)$. The
right diagram includes some small random perturbations.}
\label{Fig: Supp}
\end{figure}

We give more details on how to numerically calculate the surface states and
zero energy modes. We take a slab in the $(001)$ direction and maintain the
periodic boundary condition in the $\left( x,y\right) $-directions. Along
the $z$-direction, we work in the real space by an inverse Fourier transform of
the momentum $k_{z}$. Suppose we consider a $N_{z}$-site-thick slab, for any
fixed $(k_{x},k_{y})$, we arrange the $2N_{z}$ basis-vectors of the Hilbert
space by $(|\hspace{-0.1cm}\downarrow \rangle _{1},|\hspace{-0.1cm}\uparrow
\rangle _{1},\cdots ,|\hspace{-0.1cm}\downarrow \rangle _{N_{z}},|\hspace{%
-0.1cm}\uparrow \rangle _{N_{z}})$, where the subscript denotes the site
number. After the inverse Fourier transform, the Hamiltonian can be written
in general as $\mathscr{H}=\sum_{k_{x},k_{y}}\mathcal{H}^{k_{x},k_{y}}$,
where
\begin{equation}
\mathcal{H}^{k_{x},k_{y}}=\sum_{i=1}^{2N_{z}}%
\sum_{j=1}^{2N_{z}}t_{ij}^{k_{x},k_{y}}c_{k_{x},k_{y,}i}^{\dagger
}c_{k_{x},k_{y,}j}.  \label{eq:Hkxky}
\end{equation}%
We aim to find an analytical expression for $t_{ij}^{k_{x},k_{y}}$. From the
text, the Hamiltonian in the momentum space reads 
\begin{equation}
\mathscr{H}=\sum_{\mathbf{k}}\Psi ^{\dagger }(\mathbf{k})\mathcal{H}(\mathbf{%
k})\Psi (\mathbf{k})=\sum_{\mathbf{k}}\{u_{z}c_{\mathbf{k},\uparrow
}^{\dagger }c_{\mathbf{k},\uparrow }-u_{z}c_{\mathbf{k},\downarrow
}^{\dagger }c_{\mathbf{k},\downarrow }+[(u_{x}+iu_{y})c_{\mathbf{k},\downarrow
}^{\dagger }c_{\mathbf{k},\uparrow }+h.c.]\},
\end{equation}%
where 
\begin{eqnarray}
u_{x}+iu_{y} &=&2(\sin k_{x}+it\sin k_{y})^{p}[\sin k_{z}-i(\cos k_{x}+\cos
k_{y}+\cos k_{z}+h)]^{q}  \label{eq:uxuy} \\
u_{z} &=&(\sin ^{2}k_{x}+t^{2}\sin ^{2}k_{y})^{p}-[\sin ^{2}k_{z}+(\cos
k_{x}+\cos k_{y}+\cos k_{z}+h)^{2}]^{q}.  \label{eq:uz}
\end{eqnarray}%
Since we only perform inverse Fourier transform in the $z$ direction and keep $%
(k_{x},k_{y})$ in the momentum space, we can take $k_{x}$ and $k_{y}$ as
constants. Let $A=2(\sin k_{x}+it\sin k_{y})^{p}\;%
\text{and }B=-i(\cos k_{x}+\cos k_{y}+h)$, then Eq. (\ref{eq:uxuy}) reduces
to 
\begin{eqnarray}
u_{x}+iu_{y} &=&A(-ie^{ik_{z}}+B)^{q}  \notag \\
&=&A\sum_{\kappa =0}^{q}\left( 
\begin{array}{c}
q \\ 
\kappa 
\end{array}%
\right) (-i)^{\kappa }e^{i\kappa k_{z}}B^{q-\kappa }  \label{eq:uxuy1} \\
&=&\sum_{\kappa =0}^{q}D_{\kappa }e^{i\kappa k_{z}},  \notag
\end{eqnarray}%
where $D_{\kappa }=A\left( 
\begin{array}{c}
q \\ 
\kappa 
\end{array}%
\right) (-i)^{\kappa }B^{q-\kappa }$. Similarly, for the $u_{z}$ term, we define $%
R=(\sin ^{2}k_{x}+t^{2}\sin ^{2}k_{y})^{p}$, $S=1+(\cos k_{x}+\cos
k_{y}+h)^{2}$, $T=(\cos k_{x}+\cos k_{y}+h)$ and $Q=S/T$. Eq. (\ref{eq:uz})
then reduces to 
\begin{eqnarray*}
u_{z} &=&R-T^{q}(Q+e^{ik_{z}}+e^{-ikz})^{q} \\
&=&R-\sum_{\alpha +\beta +\kappa =q}\left( 
\begin{array}{c}
q \\ 
\alpha ,\beta ,\kappa 
\end{array}%
\right) T^{q}Q^{\kappa }e^{i(\alpha -\beta )k_{z}} \\
&=&R-\sum_{\alpha +\beta +\kappa =q}J_{\alpha \beta \kappa q}e^{i(\alpha
-\beta )k_{z}},
\end{eqnarray*}%
where $\left( 
\begin{array}{c}
q \\ 
\alpha ,\beta ,\kappa 
\end{array}%
\right) =\dfrac{q!}{\alpha !\beta !\kappa !}$ is the trinomial coefficient
and $J_{\alpha \beta \kappa q}=\left( 
\begin{array}{c}
q \\ 
\alpha ,\beta ,\kappa 
\end{array}%
\right) T^{q}Q^{\kappa }$. Now we are ready to perform the inverse Fourier
transform in the $z$ direction: 
\begin{eqnarray*}
c_{k_{x},k_{y},k_{z},\sigma } &=&\frac{1}{\sqrt{N_{z}}}%
\sum_{z}e^{izk_{z}}c_{k_{x},k_{y},z,\sigma }, \\
c_{k_{x},k_{y},k_{z},\sigma }^{\dagger } &=&\frac{1}{\sqrt{N_{z}}}%
\sum_{z}e^{-izk_{z}}c_{k_{x},k_{y},z,\sigma }^{\dagger }.
\end{eqnarray*}%
After the transformation, we obtain 
\begin{eqnarray}
\mathcal{H}^{k_{x},k_{y}} &=&[(Rc_{k_{x},k_{y},z,\uparrow }^{\dagger
}c_{k_{x},k_{y},z,\uparrow }-\sum_{\alpha +\beta +\kappa =q}J_{\alpha \beta
\kappa q}c_{k_{x},k_{y},z,\uparrow }^{\dagger }c_{k_{x},k_{y},z-\alpha
+\beta ,\uparrow })-(\uparrow \rightarrow \downarrow )]  \notag \\
&&+[\sum_{\kappa =0}^{q}D_{\kappa }c_{k_{x},k_{y},z,\downarrow }^{\dagger
}c_{k_{x},k_{y},z-\kappa ,\uparrow }+h.c.].  \label{eq:Hkxky1}
\end{eqnarray}%
Comparing Eq.(\ref{eq:Hkxky1}) with Eq. (\ref{eq:Hkxky}), we find the
expressions 
\begin{eqnarray*}
t_{2k,2l}^{k_{x},k_{y}} &=&R\delta _{k,l}-\sum_{\alpha +\beta +\kappa
=q}J_{\alpha \beta \kappa q}\delta _{2k,2l+\alpha -\beta }, \\
t_{2k-1,2l-1}^{k_{x},k_{y}} &=&-R\delta _{k,l}+\sum_{\alpha +\beta +\kappa
=q}J_{\alpha \beta \kappa q}\delta _{2k,2l+\alpha -\beta }, \\
t_{2k-1,2l}^{k_{x},k_{y}} &=&\sum_{\kappa =0}^{q}D_{\kappa }\delta
_{2k-1,2l+\kappa }, \\
t_{2k,2l-1}^{k_{x},k_{y}} &=&\sum_{\kappa =0}^{q}D_{\kappa }^{\ast }\delta
_{2k+\kappa ,2l-1},
\end{eqnarray*}%
where $0\leqslant k,l\leqslant N_{z}$. Hence, for each $k_{x}$ and $k_{y}$,
we have a $2N_{z}\times 2N_{z}$ matrix $t^{k_{x},k_{y}}$ with its $(i,j)$-th
entry $t_{ij}^{k_{x},k_{y}}$. Numerically diagonalizing this matrix for
fixed $k_{x}$ and $k_{y}$, we obtain the energy spectrum of $2N_{z}$ states.
For each $k_{x}$ and $k_{y}$, we count the number of surface states by
noticing that surface state energies have huge gaps from the bulk state
energies. Fig.\ \ref{Fig: Supp} shows the number of edge states for all $%
k_{x}$ and $k_{y}$ values by imposing a minimum relative separation from the
bulk. The number of surface states is counted as the union of all edge
states for all ($k_{x}$, $k_{y}$). The right diagram shows the case where
small random perturbations are included. We see that the surface states are robust to random perturbations without any symmetry constraint. In the text, for easy visualization, we plotted the energy spectrum of the Hamiltonian in Fig.\ 3 at fixed $%
(k_{x},k_{y})\approx (0.72,0.72)$. Each in-gap point corresponds to a
surface state.

\end{document}